\documentclass[aps,pre,twocolumn,groupedaddress]{revtex4-1}
\usepackage{graphicx}
\usepackage{dcolumn}
\usepackage{bm}
\usepackage{amsmath}
\usepackage{float}
\usepackage{subfigure}

\begin{document}


\title{Suppression of work fluctuations by optimal control: An approach based on Jarzynski's equality}


\author{Gaoyang Xiao$^1$ and Jiangbin Gong$^{2,}$}
\email[]{phygj@nus.edu.sg}
\affiliation{$^1$Department of Physics, National University of Singapore, Singapore 117542\\
$^2$Department of Physics and Centre for Computational Science and Engineering, National University of Singapore, Singapore 117542}


\date{\today}

\begin{abstract}
Understanding and manipulating work fluctuations in microscale and nanoscale systems
are of both fundamental and practical interest. For example, aspects of work fluctuations
will be an important factor in designing nanoscale heat engines.
In this work, an optimal control approach directly exploiting Jarzynski's equality is proposed to
effectively suppress the fluctuations in the work statistics,  for systems (initially at thermal equilibrium) subject to a work protocol but isolated from a bath during the protocol.  The control strategy is to
 minimize the deviations of individual values of $e^{-\beta W}$ from their ensemble average given by $e^{-\beta \Delta F}$, where $W$ is the work, $\beta$ is the inverse temperature, and $\Delta F$ is the free energy difference
between two equilibrium states.  It is further shown that even when the system Hamiltonian is not fully known, it is still possible to suppress work fluctuations through a feedback loop, by refining the control
target function on the fly through Jarzynski's equality itself.   Numerical experiments are based on linear and nonlinear parametric oscillators.  Optimal control results for linear parametric oscillators are also benchmarked with early results based on accelerated adiabatic processes.
\end{abstract}


\maketitle

\section{Introduction}
Aspects of thermodynamics at the nanoscale have attracted great attention in recent years. In a small system with few number of particles, thermal fluctuations and quantum fluctuations in work, heat, and other quantities
can be comparable to their ensemble mean values.  Understanding and manipulating these fluctuations
hence become an interesting and important topic.  From a fundamental point of view,
the connections between nonequilibrium statistical fluctuations with equilibrium properties, such as those
established through fluctuation theorems (e.g., Jarzynski equality \cite{Jarzynski.97.PRL,Jarzynski.97.PRE}
 and crooks theorem \cite{crooks}), have laid a solid and fruitful foundation for nanoscale thermodynamics \cite{Campisi.11.RMP}.  It is interesting to note that the existence of such type of fluctuation theorems does not rule out the possibility to control statistical fluctuations. That is, under appropriate conditions, it is still possible
to manipulate nonequilibrium processes and alter the statistical fluctuations in thermodynamic quantities, without
violating any fundamental law in thermodynamics or modifying the fluctuation theorems themselves.  In the case of Jarzynski's equality, i.e., $\langle e^{-\beta W}\rangle=e^{-\beta \Delta F}$ \cite{Jarzynski.97.PRL,Jarzynski.97.PRE}, where $W$ is work done on a system initially at thermal equilibrium with inverse temperature $\beta$ and $\Delta F$ is the free energy difference between the final and initial equilibrium states of the same $\beta$,
a change in fluctuations of $W$ can have an impact on how fast the statistics of $e^{-\beta W}$
may converge towards $e^{-\beta \Delta F}$.   This is the first motivation of this study.

Advances in our knowledge of thermodynamics of small systems have stimulated studies of efficient energy devices (classical or quantum) at microscale and nanoscale, sometimes consisting of few particles or few degrees of freedom.
 Of particular interest here is that several designs of microscale or nanoscale heat engines have been proposed theoretically \cite{Abah.12.PRL,Bergenfeldt.14.PRL,Zhang.14.PRL}.  For this type of applications where heat-to-work conversion efficiency and the power of work output are apparently crucial, the stability or reliability of work output also becomes an important performance indicator \cite{Deng.13.PRE,dario}.  This is  because the cycle-to-cycle fluctuations in the work output  are an unavoidable characteristic of nanoscale heat engines.   Given two heat engines with the same mean work output per cycle, it seems desirable to prefer the one with less fluctuations and hence more uniform output.     This understanding further motivates
  us to ask the following question: how to systematically suppress work fluctuations in a given protocol (such as in one step of a heat engine cycle that does not involve any heat exchange with a bath)?

It has been realized that work fluctuations in some controlled processes can be relatively smaller than those in bare uncontrolled processes.
 In particular, work fluctuations in an adiabatic process (here ``adiabatic" means very slow, as compared with the system's own natural time scale) are relatively small \cite{Deng.13.PRE}. However, an adiabatic process in the standard sense is simply too slow, and as such
the power of work output based on adiabatic processes vanishes. For this reason it is natural to
consider accelerated adiabatic processes to suppress work fluctuations and increase the power of work output at the same time~\cite{Deng.13.PRE,campoarxiv}.  Certainly, accelerated adiabatic processes just represent
 a special type of controlled dynamics, and extensions based on alternative constructions of a control field should be possible.  This paper is precisely one extension of Ref.~\cite{Deng.13.PRE} in order to suppress  work fluctuations in a wider variety of systems. The extension is based on an optimal control framework \cite{Shi.90.JCP,Peirce.88.PRA,Shi.91.CPC,Sundermann.99.JCP,Ohtsuki.04.JCP} that is far more versatile and far more flexible. We shall also use known results based on accelerated adiabatic processes to benchmark our optimal control results.

To see the necessity and advantages of an optimal-control-based extension, let us first
make brief comments on early studies of accelerated adiabatic processes \cite{Demirplak.08.JCP,Demirplak.05.JPCB,Demirplak.03.JPCA,Berry.09.JPAMT,Masuda.10.PRSLSA,Chen.10.PRL,Chen.10.PRLa,Choi.11.PRA,Schaff.11.NJP,Bason.12.NP,Zhang.13.PRL,Jarzynski.13.PRA,Campo.11.PRA}. For the sake of discussions later on, we divide accelerated adiabatic processes known to date into two types (though these two types can even be regarded as being equivalent upon a transformation ~\cite{Deffner.14.PRX} and there is no clear distinction in the literature). In the first type, an additional control Hamiltonian is introduced to drive a system (within a short time scale), such that the evolution of the system, either classical or quantum, still follows the adiabatic evolution of the original bare system. Here we call this type of control as fast-forward adiabatic driving (FFAD) \cite{Masuda.10.PRSLSA}.
The extra control Hamiltonian in realizing FFAD may be found for very simple systems but in general, its analytical form is not available and and it is
also challenging to precisely implement it experimentally. The second type of accelerated adiabatic processes, which we call ``shortcuts to adiabaticity" (STA) \cite{mugareview}, relies on special time dependence of system parameters. In this sense, the required form of the control Hamiltonian is already contained in the original bare Hamiltonian.  Within a finite time, the system under STA control can  reach the same final states as those reached in conventional adiabatic processes. However, during an STA process the system may make nonadiabatic transitions \cite{Tu.14.PRE,campoarxiv}.  The realization of STA has been established in several simple systems such as two-level systems or linear parametric oscillators. As  Ref.~\cite{Deffner.14.PRX} showed, STA in known examples can be understood as a result of a canonical transformation of FFAD (or vice versa). Though Campo and Jarzynski extended STA to more complicated systems such as many-body systems \cite{Campo.13.PRL}, a general discussion by them also indicates that this method can only apply to systems with the so-called ``scale-invariant driving" \cite{Deffner.14.PRX}.  For these reasons both FFAD and STA require full knowledge of the system Hamiltonian, and this requirement
presents a limitation when we attempt to use accelerated adiabatic processes to suppress work fluctuations in general situations.

In this work we restrict ourselves to classical systems (but our optimal control approach can be easily extended to quantum systems). As such only work fluctuations across a thermal ensemble
are to be suppressed.   Our classical results should be also useful in guiding possible manipulations of work fluctuations in quantum systems, especially in cases with a relatively high temperature.
Specifically,  we develop a method to design control fields to suppress work fluctuations based on the well-known optimal control theory (OCT). The peculiarities of our OCT here lie in two aspects. First,
we need to handle a thermal ensemble in order to consider work fluctuations. Thus our control is ensemble-based optimal control. Second, we need to design a useful control target function in order to reach our goal. As shown below, our control target function directly exploits Jarzynski's equality, and the constructed target function is biased against deviations of individual values of $e^{-\beta W}$ from their ensemble average, i.e., $e^{-\beta \Delta F}$.
We show that our optimal control method built in this manner can be applied to a wide variety of systems, including highly nonlinear systems. When it is applied to previously known simple systems, it has essentially the same quantitative performance as that based on accelerated adiabatic processes, but now the control fields have many possible forms. Furthermore, our OCT method can be extended to implement a feedback mechanism, so as to  suppress work fluctuations in systems with unknown system parameters. This is done by guessing the control target function iteratively through Jarzynski's equality again.

This paper is arranged as follow. In Sec.~II, we apply OCT to the topic of suppression of classical work fluctuations, with necessary details. In Sec.~III, we test our approach using a simple system and compare our results with those obtained based on accelerated adiabatic processes. In
Sec.~IV, we consider a nonlinear oscillator and show that our OCT approach can still operate well.  In Sec.~V, we consider a nonlinear oscillator with some unknown system parameters.  We show that, in the spirit of feedback optimal control, it is possible to refine the control target function on the fly, and then in the end  it is still possible to find a control field to suppress the work fluctuations.  The last section of this paper gives a brief summary.  An appendix regarding numerical implementation of our OCT is also given.

\section{Theory of Optimal Control of work fluctuations: use of Jarzynski's equality to construct a control target function }
Let us consider a general time-dependent classical system with Hamiltonian $H_0(p,q,\alpha(t))$, where $(p,q)$ represents phase space coordinates (we assume one-dimensional problems), and $\alpha(t)$ is a time-dependent parameter due to a work protocol starting from $t=0$ to $t=\tau$.  For example, this $\alpha$ can represent the frequency of a parametric oscillator, the length of a pendulum, or the width of a box, etc.
Such a classical system is initially prepared at thermal equilibrium, with the phase space probability density $\rho$ given by
\begin{equation}\label{Gibbs}
  \rho(p_0,q_0,0)=\frac{1}{Z_0}\exp\{-\beta H_0[p_0,q_0,\alpha(0)]\},
\end{equation}
where $Z_0$ is the partition function associated with $\alpha=\alpha(0)$.
To suppress work fluctuations in this protocol defined by $\alpha(t)$, an extra control field is applied. Then the total  Hamiltonian $H$ of the system becomes
\begin{equation}\label{Hamlitonian}
  H(p,q,t)=H_0[p,q,\alpha(t)]+H_c[p,q,A(t)],
\end{equation}
where $H_c$ is the control Hamiltonian and $A(t)$ is assumed to be the time-dependent amplitude of a control field. The evolution of $(p,q)$ obeys the following Hamilton's equation of motion:
       $\dot{q}   = \frac{\partial H}{\partial p}$ and
       $\dot{p}   = -\frac{\partial H}{\partial q}$.

Next we consider a certain physical quantity $Q[p(t),q(t),t]$ evaluated at $t=\tau$, which is written as $Q(p_\tau,q_\tau,\tau)$. Consistent with this notation and for convenience, we also define $p_0=p(0)$ and $q_0=q(0)$ from now on.
The minimization of the ensemble-averaged value of $Q[p_\tau,q_\tau,\tau]$ will be important in our OCT.
The thermal ensemble average of $Q[p_\tau, q_\tau,\tau]$ is given by
\begin{equation}\label{L1}
  L_1=\langle Q \rangle \equiv \frac {1}{Z_0} \int_\Gamma e^{-\beta H_0[p_0,q_0,\alpha(0)]} Q(p_\tau,q_\tau,\tau)dp_0dq_0,
\end{equation}
where $\Gamma$ represents the entire phase space.
We call $L_1$ defined above a target function since it will be a quantity we hope to minimize.
For a control problem, typically a cost function is also needed to reflect a cost-related constraint. Here we define the cost function using the amplitude of the control field, i.e.,
\begin{equation}\label{L2}
  L_2=\frac{1}{2}\int_{0} ^\tau \kappa A^{2}(t)dt,
\end{equation}
where $\kappa$ is a weightage factor that may depend on time.  The larger $\kappa$ is, the more constraints posted on the control field due to cost considerations.
The overall target function can then be defined as $J=L_1+L_2$.
That is, the problem becomes to minimize $J$ under the general dynamical constraints reflected by Hamilton's equations of motion.  To that end we introduce two Lagrange multiplier as functions of $(p,q)$, denoted by $\lambda(q,p,t)$ and $l(q,p,t)$.  We then minimize $\bar{J}$ instead, with
\begin{widetext}
\begin{equation}\label{Jbar}
  \bar{J}=L_1+L_2+\int_\Gamma \int_{0} ^\tau \lambda(q,p,t){\,}(\dot{q}-\frac{\partial {H}}{\partial {p}})\ dq_0dp_0dt+\int_\Gamma \int_0 ^\tau l(q,p,t){\,} (\dot{p}+\frac{\partial{H}}{\partial{q}})\ dq_0dp_0dt.
\end{equation}
\end{widetext}
Let $(\delta p,\delta q)$ be the variation in $(p,q)$ due to $\delta A(t)$, an arbitrary variation in $A(t)$, . Then the variation in $\bar{J}$ due to $\delta A(t)$ is found to be
\begin{widetext}
\begin{equation}\label{deltaJ}
  \begin{split}
    \delta {\bar{J}} & = \frac{1}{Z_0} {\int_\Gamma} e^{-{\beta} H_0(q_0,p_0,0)}{\,}\left(\frac{\partial Q}{\partial q_\tau}{\delta q_\tau}+\frac{\partial Q}{\partial p_\tau}{\delta p_\tau}\right)dq_0dp_0+{\int_\Gamma} ({\lambda_\tau} {\delta q_\tau} + {l_\tau} {\delta p_\tau}) dq_0 dp_0\\
      &-{\int_\Gamma}{\int_0 ^\tau} {{\lambda(q,p,t)}{\,}\left({\frac{\partial^{2}{H}}{\partial{p^2}}}{\delta p}+{\frac{\partial^{2}{H}}{\partial{p} \partial {q}}}{\delta q}\right)dq_0 dp_0dt}\\
      &+{\int_\Gamma}{\int_0 ^\tau} {{l(q,p,t)}{\,}\left({\frac{\partial^{2}{H}}{\partial{q^2}}}{\delta q}+{\frac{\partial^{2}{H}}{\partial{p} \partial {q}}}{\delta p}\right)dq_0 dp_0dt}\\
      &-{\int_\Gamma} {\int_0 ^\tau} [\dot{\lambda}(q,p,t){\delta q}+\dot{\l}(q,p,t){\delta p}]dq_0dp_0dt\\
      &+{\int_\Gamma} {\int_0 ^\tau} \left[l(q,p,t){\frac{\partial^{2}{H}}{\partial{q} \partial {A}}}-\lambda(q,p,t){\frac{\partial^{2}{H}}{\partial{p} \partial {A}}}\right]\delta A{\,}dq_0dp_0dt+{\int_0 ^\tau}\kappa A(t)\delta A{\,}dt.
  \end{split}
\end{equation}
\end{widetext}
To minimize $\bar{J}$ we let $\delta J=0$. Since the variation is arbitrary,
one has the following relations:
\begin{equation}\label{itteration}
  \left\{
  \begin{split}
      & \frac{e^{-{\beta} H_0(q_0,p_0,0)}}{Z_0}{\,} \frac{\partial Q}{\partial q_\tau}+{\lambda_\tau}=0, \\
      & \frac{e^{-{\beta} H_0(q_0,p_0,0)}}{Z_0}{\,} \frac{\partial Q}{\partial p_\tau}+{l_\tau}=0, \\
      & -\dot{\lambda}(q,p,t)+l(q,p,t) {\frac{\partial^{2}{H}}{\partial{q^2}}}-{\lambda(q,p,t)} {\frac{\partial^{2}{H}}{\partial{p} \partial {q}}}=0,\\
      & -\dot{\l}(q,p,t)+l(q,p,t) {\frac{\partial^{2}{H}}{\partial{p} \partial {q}}}-{\lambda(q,p,t)} {\frac{\partial^{2}{H}}{\partial{p^2}}}=0,\\
      & {\int_\Gamma} \left[l(q,p,t){\frac{\partial^{2}{H}}{\partial{q} \partial {A}}}-\lambda(q,p,t){\frac{\partial^{2}{H}}{\partial{p} \partial {A}}}\right]dq_0dp_0+\kappa A(t)=0.
  \end{split}
  \right.
\end{equation}
The list of relations in Eq.~(\ref{itteration}) can be numerically solved by an iteration procedure and some details are presented in Appendix.

After outlining the general steps in OCT, we turn to our main objective, which is to minimize the statistical work fluctuations if a work protocol is applied to a thermal ensemble. For the above-defined protocol of $\alpha(t)$ during which there is no interaction with a heat bath,  the inclusive {\it work} as a function of $(p_0,q_0)$ is given by \cite{Jarzynski.97.PRL,Jarzynski.07.CRP,Campisi.11.RMP}:
\begin{equation}\label{Work}
  W_{\tau}(q_0,p_0)=H[q(q_0,p_0,\tau),p(q_0,p_0,\tau),\tau]-H(q_0,p_0,0),
\end{equation}
where $q(q_0,p_0,\tau),p(q_0,p_0,\tau))$ is simply the phase space coordinates at the end of the protocol.
The ensemble average of $W$ is
\begin{widetext}
\begin{equation}\label{averageW}
  \langle W \rangle=\int_\Gamma \frac{e^{-\beta H(q_0,p_0,0)}}{Z_0}\{H[q(q_0,p_0,\tau),p(q_0,p_0,\tau),\tau]-H(q_0,p_0,0)\}dq_0dp_0.
\end{equation}
\end{widetext}
In order to suppress work fluctuations across the thermal ensemble, one may now choose an appropriate target function to minimize via OCT. A naive and simple choice of the target function can be just the variance squared of $W$. That is, one may choose the ensemble average
$\langle (W-\langle W \rangle)^2 \rangle$ as the target function.
However, it turns out that, different control fields can yield different values of $\langle W \rangle$, and as a result one does not know $\langle W \rangle$ beforehand.  Thus such type of intuitive  choice of the control target function is not adopted here.  At the point,  Jarzynski's equality comes as a rescue.
Note first that
\begin{equation}
  \langle e^{-\beta W} \rangle  = e^{-\beta \tilde{\Delta} F},
\end{equation}
where $\tilde{\Delta} F$ refers to the free energy difference associated with the total Hamiltonian $H=H_0+H_c$.
So long as the initial and final values of $H_c$ are zero (this is easily realized in the next section)
, then $\tilde{\Delta} F$ will be the same as ${\Delta} F$, the free energy difference associated with $H_0[p,q,\alpha(0)]$ and $H_0[p,q,\alpha(\tau)]$.
This being the case, no matter what the to-be-found control field is, the ensemble-averaged value of $e^{-\beta W}$ is fixed, which is given by $e^{-\beta {\Delta} F}$.  We are thus motivated to design a control target function as follows:
\begin{equation}\label{fluctuation}
  \begin{split}
    L_1 & = \left\langle \left[e^{-\beta W}-\langle e^{-\beta W} \rangle\right]^2 \right\rangle\\
      & = {\int_\Gamma} \frac{e^{-{\beta} H_0(q_0,p_0,0)}}{Z_0} (e^{-\beta W}-e^{-\beta \Delta F})^2 dq_0dp_0.
  \end{split}
\end{equation}
The above-defined form of the control target function $L_1$ is intriguing,
because minimization of this function
is then to directly suppress the derivations of possible individual values of $e^{-\beta W}$ from their ensemble-average value predicted by Jarzynski's equality. In this regard, our OCT framework directly exploits Jarzynski's equality. It is also expected that the found optimal control field can remove those rare trajectories with rare values of $e^{-\beta W}$, which slow down the convergence of simulation results towards Jarzynski's equality.

\section{Optimal Control of Work Fluctuations in Linear Parametric Oscillators}
As a benchmark step, in this section we
consider parametric (linear) oscillator systems.  In particular, for such systems, the control Hamiltonian to realize accelerated adiabatic processes can be found easily, for both scenarios of
fast-forward adiabatic driving (FFAD) and shortcuts to adiabaticity (STA). We can hence compare
the performance of our optimal control fields with those found in FFAD and STA.

The Hamiltonian of a parametric oscillator with a time-dependent frequency, all in dimensionless units, is described by
\begin{equation}\label{Harmonic}
  H_0=\frac{p^2}{2m}+\frac{1}{2} m\omega^2(t) q^2.
\end{equation}
For FFAD, the extra control field is \cite{Deng.13.PRE,Jarzynski.13.PRA}
\begin{equation}\label{FFA}
  H_c^{\text{FFAD}}(t)=-\frac{\dot{\omega}}{2\omega}\,pq.
\end{equation}
For STA, the control field is found to be \cite{Campo.13.PRL}
\begin{equation}\label{STA}
  H_c^{\text{STA}}(t)=\frac{1}{2} \left[-\frac{3\dot{\omega}^2}{4\omega ^2}+\frac{\ddot{\omega}}{2\omega}\right] q^2.
\end{equation}
As an example, we choose a frequency protocol \cite{Campo.13.PRL}
\begin{eqnarray}\label{frequency}
  \omega(t)&=&\omega_0+10(\omega_\tau-\omega_0){\,}\left(\frac{t}{\tau}\right)^3
  \nonumber \\ && -15(\omega_\tau-\omega_0){\,}\left(\frac{t}{\tau}\right)^4+6(\omega_\tau-\omega_0){\,}\left(\frac{t}{\tau}\right)^5
\end{eqnarray}
such that the above two control fields are indeed zero at $t=0$ or $t=\tau$.
Besides, note that
\begin{equation}\label{increasew}
    \dot{\omega} = 30(\omega_\tau-\omega_0)\left(\frac{t}{\tau}\right)^2\frac{1}{\tau}\left(1-\frac{t}{\tau}\right)^2
\end{equation}
So $\dot{\omega}\geq0$ during the entire protocol if we choose $\omega_\tau>\omega_0$. In our calculations we
choose $\omega_0=10.0$ and $\omega_\tau=10\sqrt{3}$ in dimensionless units. Then, it is easy to show that
\begin{equation}\label{positiveW}
   W_{\tau}=\int_0 ^\tau \frac{\partial H}{\partial \omega}\;\dot{\omega}dt\geq0.
\end{equation}
That is, along an arbitrary trajectory, the work value is always positive.
We further set $\tau$ to be as small as $0.001$ (as compared with $1/\omega_0$),
such that the process will be highly nonadiabatic were there no control fields.

To benchmark our OCT approach outlined above with FFAD and STA, we introduce two kinds of control fields, i.e., $A(t)pq$ and $\frac{1}{2}A(t)q^2$ (with $A(t)$ to be found numerically), in parallel with the above two control fields $H_c^{\text{FFAD}}(t)$ and $H_c^{\text{SAT}}(t)$ to realize FFAD and STA.
The weightage factor $\kappa$ in Eq. (\ref{L2}) is set to be small, since here we are not much concerned with the cost of the control field. Nevertheless, the time-dependence of $\kappa$ can be designed to further alter the profile of the field amplitude.  Here we propose to use $\kappa(t)={\tilde{\kappa}}/{\sin {\frac{\pi t}{\tau}}}$, where $\tilde{\kappa}$ is a small constant.
Such time-dependence of $\kappa$ makes $\kappa$ diverge at $t=0$ and $t=\tau$, and as a consequence
the numerically found $A(t)$ will become zero automatically at these two boundary times due to the cost function $(\ref{L2})$ in the OCT.

\begin{figure}[H]
\centering
\subfigure[]{
\label{Fig1a}
\includegraphics[width=8.4cm]{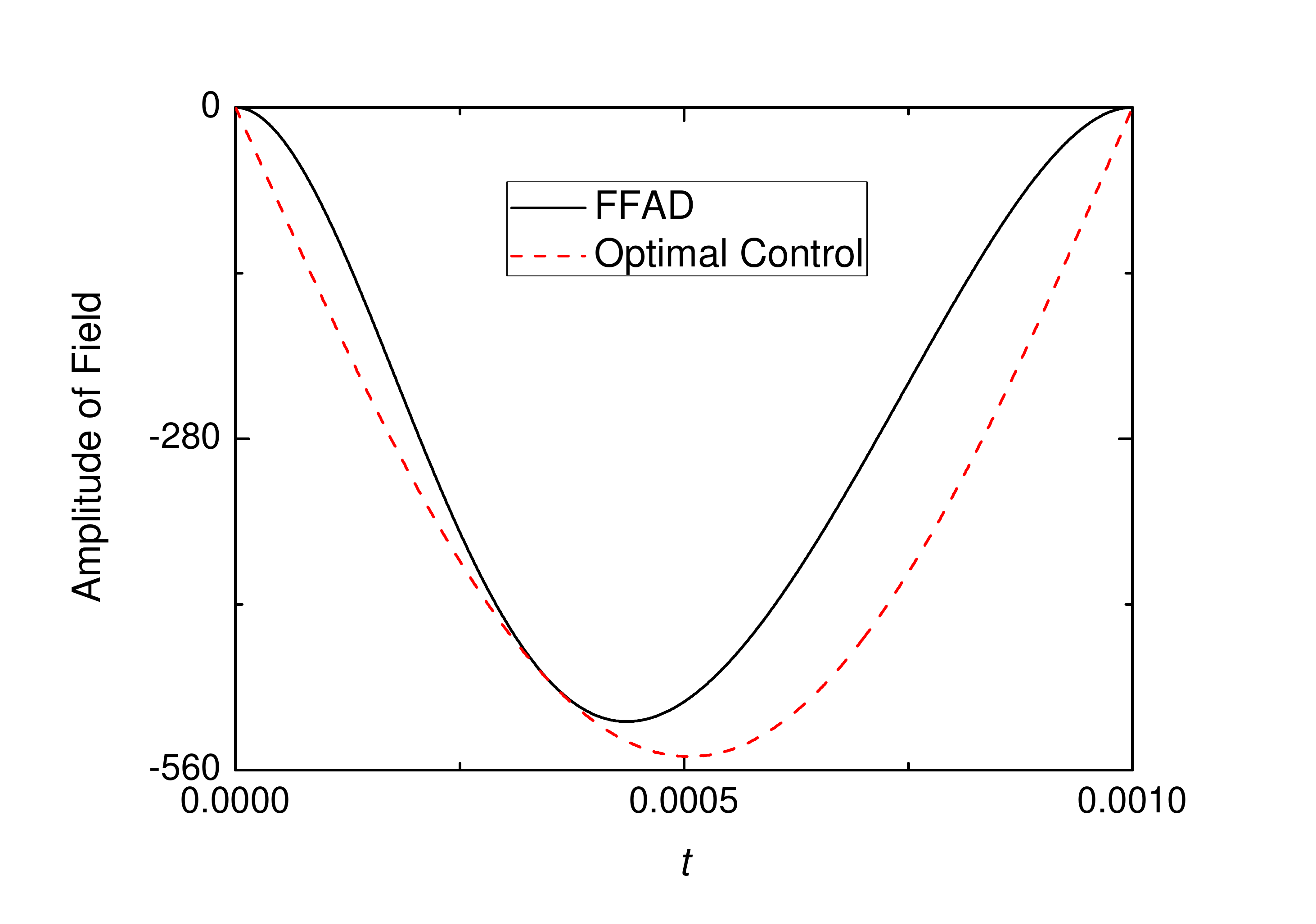}}
\subfigure[]{
\label{Fig2a}
\includegraphics[width=8.4cm]{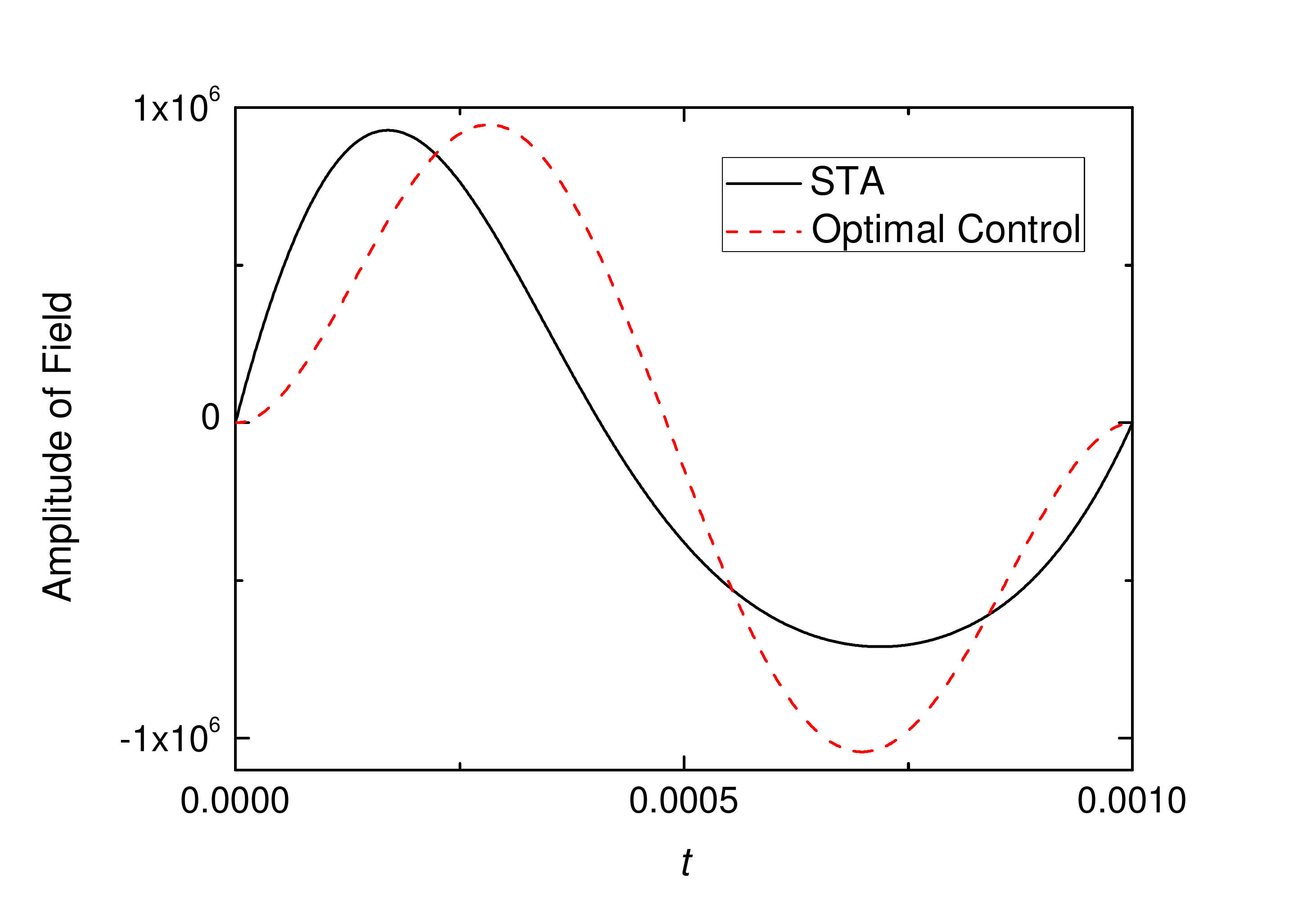}}
\caption{(color online) Time dependence of optimal control field for a parametric
 linear oscillator whose frequency time dependence is given in Eq.~(\ref{frequency}), with
$\omega(0)=10$, $\omega(\tau)=10\sqrt{3}$, and $\tau=0.001$.  All the plotted quantities here
 and in other figures are in scaled and hence dimensionless units. The inverse temperature is set to be
 $\beta=1.0$.
(a) The control field amplitude $A(t)$ of $pq$-type optimal control as compared with that
 based on fast-forward adiabatic driving (FFAD).
 (b) The control field amplitude $A(t)$ of $q^2$-type optimal control
  as compared with that in the shortcuts to adiabaticity (STA) approach.
  Note that the field amplitudes in (b) are much higher than those in (a).}
\label{Fig1}
\end{figure}

With the form of the control field and the weightage factor $\kappa$ in OCT both specified, numerical iterations based on  Eq.~(\ref{itteration}) can then yield explicit solutions of $A(t)$.  The results are
shown in Fig.~\ref{Fig1} for both $pq$-type and $q^2$-type optimal control.
Their respective time dependence is also compared with the
corresponding results to realize FFAD and STA.  Note that in all these cases the control field amplitude is zero in the beginning or at the end. It is also seen that $A(t)$ found from our  optimal control
is different from those in FFAD or STA, though the difference for the shown computational example is not yet dramatic.

\begin{figure}
  \centering
  \includegraphics[width=8.4cm]{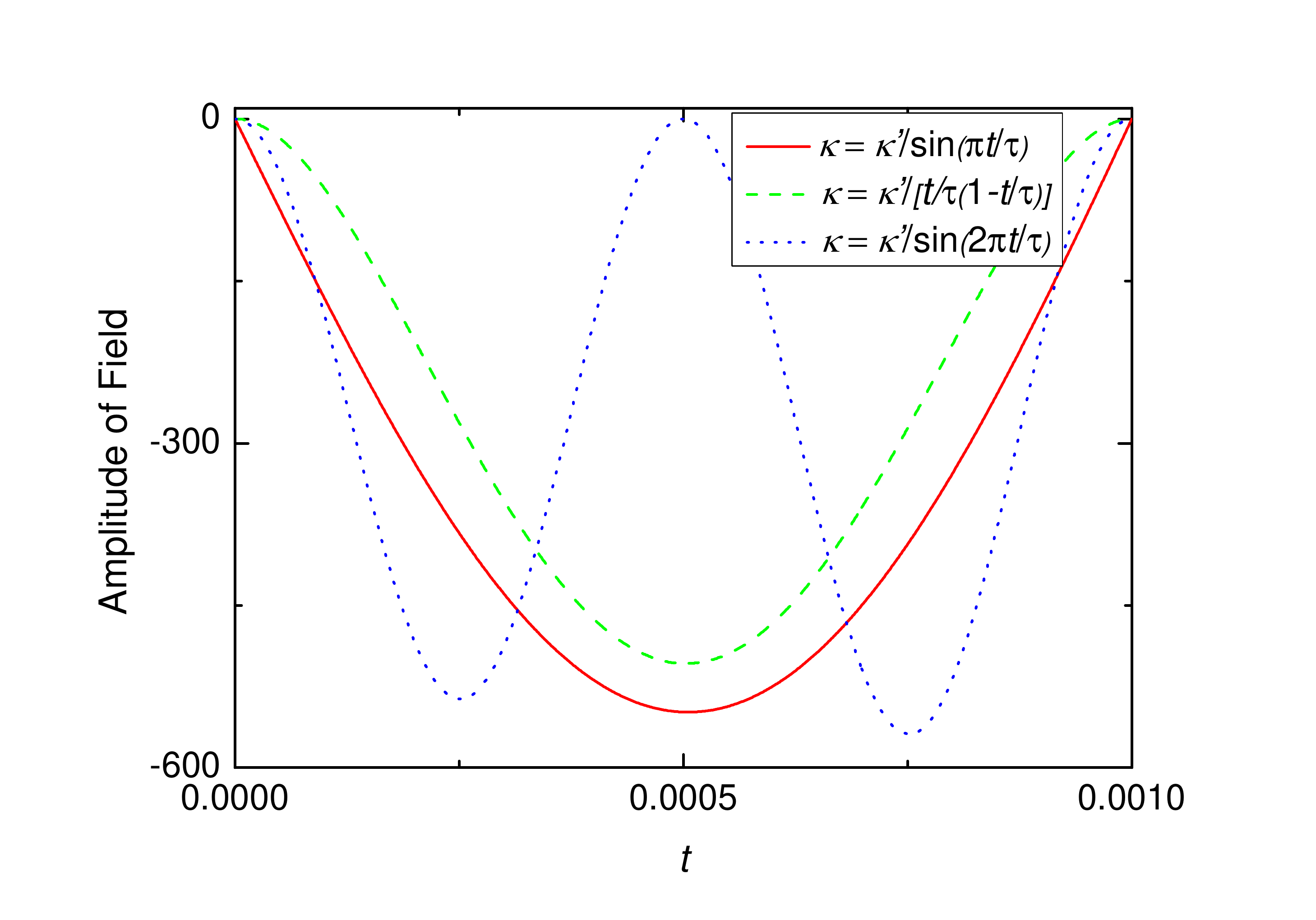}\\
  \caption{(color online) Time dependence of $pq$-type optimal control field, choosing
   different weightage function $\kappa(t)$ (indicated on the panel)
    in the cost function defined in Eq.~(\ref{L2}).
    The system considered here and other system parameters
    are the same as in Fig.~1.  The three choices of $\kappa(t)$ lead to three different
     control fields, but all of them yield $\delta(e^{-\beta W})=0.2691$ from $10^6$ trajectories.}\label{Fig7}
\end{figure}

Next we attempt to quantitatively characterize the performance in suppressing the work fluctuations, for our OCT approach along with FFAD and STA.  In particular,
we randomly sample initial phase space points (through a standard importance sampling prcedure) according to the initial thermal probability distribution and then evolve them under the total Hamiltonian $H=H_0+H_c$. Individual values of $W$ are denoted $W_i$, and the fluctuations in $ W$ and in $ e^{-\beta W}$ are characterized by
\begin{eqnarray}\label{trajectory}
     \sigma(W) & =&  \sqrt{\frac{1}{N}\sum_{i=1} ^N (W_i-\langle W \rangle)^2}\ , \nonumber \\
     \sigma(e^{-\beta W})& = &\sqrt{\frac{1}{N}\sum_{i=1} ^N \left(e^{-\beta W_i}-\langle e^{-\beta W} \rangle\right)^2}\ ,
\end{eqnarray}
where the total number of simulation trajectories is chosen to be $N=10^6$.
The variance in both $ e^{-\beta W}$ and in $W$ itself under different control schemes are all
 presented in Table~\ref{performance1}.  A few interesting observations are in order.  First,
within expected statistical error due to a finite $N$, the bare system, FFAD, STA, OCT with $pq$-type field, and OCT with $q^2$-type field all yield the same $\langle e^{-\beta W}\rangle$. Second, though the found time dependence $A(t)$ from OCT is different from FFAD or STA, the variances in $e^{-\beta W}$ and in $W$
 obtained from OCT with $pq$-type field are all the same as those obtained in FFAD and STA. This confirms that our OCT framework is doing an excellent job in suppressing work fluctuations, to the same degree as accelerated adiabatic processes can reach. This also hints that in the parametric oscillator example here,  accelerated adiabatic processes already reach an optimized suppression of work fluctuations.
 Third, even the mean work $\langle W\rangle$ from $pq$-type OCT agrees with those obtained from FFAD and STA.
Fourth, the two variances shown in the last row of Table~\ref{performance1}, which is for OCT with $q^2$-type control, are however slightly above those in other cases.  This is because for $q^2$-type control, as also shown in Fig.~1, the required amplitude of the control field is very high, so a small weightage factor $\kappa$ chosen for the cost function can still cause a minor difference. This is also manifested $\langle W\rangle$ in the last row of Table~\ref{performance1}, which is again relatively higher than those obtained in OCT with $pq$-type field, in FFAD, or in STA.

.
\begin{table*}[htb]
\caption{The performance of suppressing work fluctuations in the absence or presence of
 several different control fields, mainly characterized by the variance in work $W$ and
   the variance in $e^{\beta W}$, using $10^6$ trajectories.  The system is a parametric linear oscillator
   whose frequency time dependence is given in Eq.~(\ref{frequency}, with $\omega(0)=10$, $\omega(\tau)=10\sqrt{3}$, and $\tau=0.001$ (duration of the protocol). The inverse temperature is set to be $\beta=1$. Note that the obtained values of $\langle e^{-\beta W} \rangle$ are all around the theoretical value $\frac{1}{\sqrt{3}}\approx0.5774$ theoretically obtained from Jarzyski's equality.}
\vskip 1em
\centering
\begin{tabular}{ccccc}\hline\hline
{\quad}Process{\quad} & ${\quad}\langle e^{-\beta W} \rangle{\quad}$ & ${\quad}\sigma (e^{-\beta W}){\quad}$ & ${\quad}\langle W \rangle{\quad}$ & ${\quad}\sigma (W){\quad}$ \\[0.5ex]
\hline
bare{\;}system & $0.5773$ & $0.3373$ & $0.9990$ & $1.4120$\\
\hline
FFAD & $0.5775$ & $0.2691$ & $0.7314$ & $0.7312$ \\
\hline
STA & $0.5775$ & $0.2691$ & $0.7314$ & $0.7312$ \\
\hline
optimal{\;}control{\;}of{\;}$pq$-type & $0.5775$ & $0.2691$ & $0.7314$ & $0.7312$ \\
\hline
optimal{\;}control of{\;}of{\;}$q^2$-type & $0.5775$ & $0.2697$ & $0.7340$ & $0.7398$ \\
\hline\hline
\end{tabular}
\label{performance1}
\end{table*}
Having successfully benchmarked our OCT approach, we now shed light on the flexibility of our optimal control approach. First of all, we can introduce different time dependence to the weightage factor $\kappa$ when accounting for the cost of the control field so as to obtain different $A(t)$, the time dependence of a control field.  For example, instead of $\kappa=\tilde{\kappa}/{\sin{\frac{\pi t}{\tau}}}$, we have considered
\begin{eqnarray}
\kappa(t)&=&\kappa_1(t)=\frac{\tilde{\kappa}}{\sin(2\pi t/\tau)}; \nonumber \\ \kappa(t)&=&\kappa_2(t)=\frac{\tilde{\kappa}}{(1-t/\tau)t/\tau}
 \end{eqnarray}
 for our $pq$-type optimal control. The obtained $A(t)$ is presented in Fig.~\ref{Fig7} in comparison with our previous result. Interestingly, although the time-dependence of the control field varies significantly with changes in the cost function weightage factor $\kappa(t)$,  all the three cases shown
 in Fig.~\ref{Fig7} yield the same variance of $e^{-\beta W}$, i.e., $\sigma(e^{-\beta W})=0.2691$.
 This is a clear demonstration that there are many possible solutions in suppressing work fluctuations to a certain level.

\begin{figure}
  \centering
  \includegraphics[width=8.4cm]{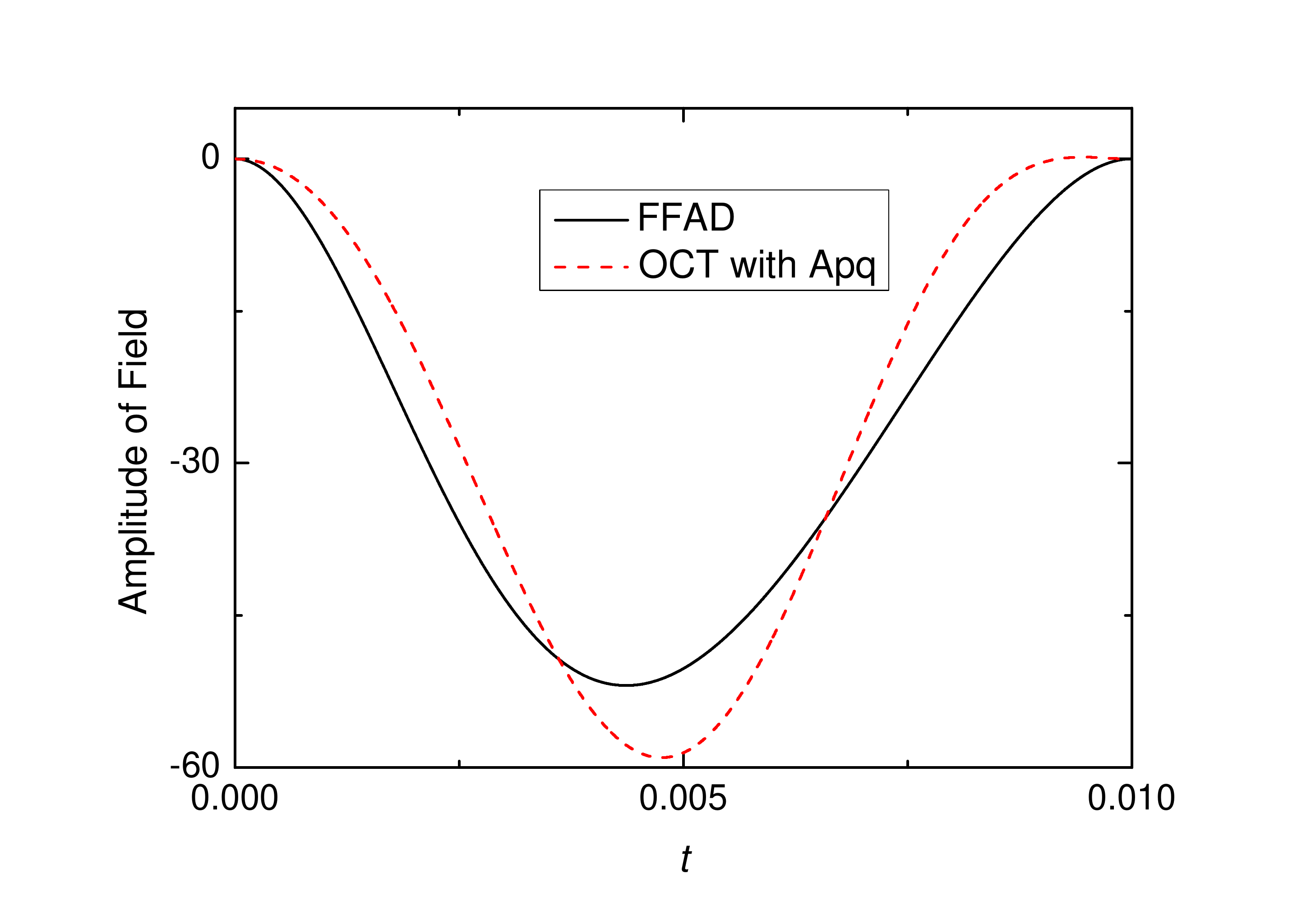}\\
  \caption{(color online) Time dependence of a $pq$-type optimal control field,
   as compared with that in FFAD, for a parametric linear oscillator.
   All system parameters are the same as in Fig.~1,
   except that $\omega(0)=100$, $\omega(\tau)=100\sqrt{3}$, and $\tau=0.01$.}\label{Fig9}
\end{figure}

 So far we have chosen $\omega\tau=0.01$, which is more or less to simulate an instantaneous limit
 for the bare system. That is,  the frequency of the parameter oscillator
 is changed rapidly as compared with the system's own time scale.  As shown above, in such a parameter regime our optimal control can suppress work fluctuations to the same degree as that achieved in accelerated
 adiabatic processes.
 To further check the usefulness of OCT, we now consider a slower protocol in which $\omega_0=100$ and $\tau=0.01$.  In this case, during the protocol the system's own bare Hamiltonian will be important in the time evolution. Our optimal control fields are found to perform also very well in this regime.  In particular,
  $\sigma(e^{-\beta W})$ without a control field decreases to $0.3212$. This is because
  nonadiabatic effects and hence work fluctuations are weaker for a slower protocol.
  Interestingly,  with FFAD or optimal control, we still find $\sigma(e^{-\beta W})=0.2691$.
  One example of $A(t)$ found from our $pq$-type optimal control is presented in Fig.~\ref{Fig9} in comparison with the field for FFAD.  As seen from Fig.~\ref{Fig9}, for a slow protocol here the required control fields to suppress fluctuations are much weaker than that presented in Fig.~1


\section{Optimal control of work fluctuations in nonlinear oscillators}
We are now ready to apply our OCT approach to systems of nonlinear oscillators under certain protocols.
Explicit solutions to realize FFAD and STA cannot be found for general nonlinear systems (excluding scale-invariant systems), but our OCT framework equally applies.  In particular, let us consider the following system Hamiltonian with a time-dependent nonlinear term:
\begin{equation}\label{anharmonic}
  H_0(t)=\frac{p^2}{2m}+\frac{1}{2} m\omega(t)^2 q^2+\varepsilon q^4 \sin\left(\pi \frac{t}{\tau}\right).
\end{equation}
where $\omega(t)$ is still the time-varying system parameter to model the same protocol defined in Eq.~(\ref{frequency}). In this system, the
 nonlinear term vanishes at $t=0$.  This choice is just for computational convenience when we sample the initial states from a thermal ensemble ensemble (which is still Gaussian). Below
 only $pq$-type optimal control are presented (note that, as observed earlier, the required
 field strength for $q^2$-type control is much larger).
In addition, for the sake of comparison,  we also examine the parallel performance of the previous FFAD field obtained in the {\it absence} of the nonlinear term.
To stress that the previous FFAD will no longer strictly give rise to accelerated adiabatic processes here due to the nonlinear term in the Hamiltonian, we call such a control approximate FFAD.

As the first computational example, we set $\beta=0.1$ and $\varepsilon=1000$, $\omega_0=10.0$, $\tau=0.001$, and then
numerically compute the fluctuations of $ e^{-\beta W}$. For approximate FFAD, we have $\sigma (e^{-\beta W})=0.2692$; whereas in our optimal control we have $\sigma (e^{-\beta W})=0.2691$.  Since these results are almost identical with our previous result in the absence of a nonlinear term, there must be a reason.
 The reason is simple. When the temperature is not high enough, the system is largely confined
to a small neighborhood of the lowest energy state and hence the nonlinear term only plays a negligible role.
 This is consistent with a recent experiment \cite{Rohringer.13.a1}, which showed that approximate FFAD can perform well in the presence of some degree of nonlinearity.  For the same value of $\beta$ and $\epsilon$, we have also considered a slow protocol with $\omega_0=100$ and $\tau=0.01$. In this case both optimal control and approximate FFAD still yield $\sigma (e^{-\beta W})=0.2691$.

 \begin{figure}
  \centering
  \includegraphics[width=8.4cm]{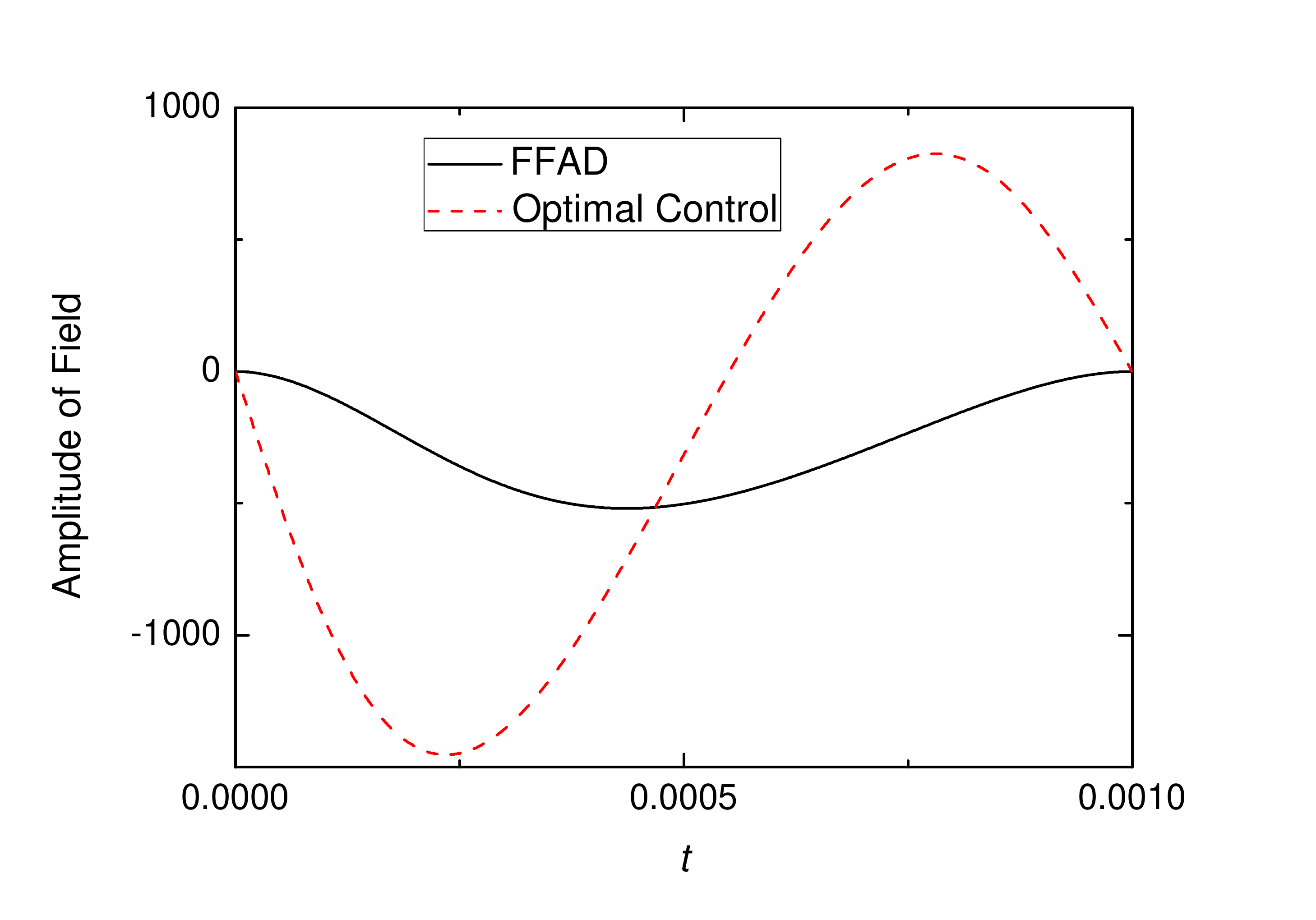}\\
  \caption{(color online) Time dependence of a $pq$-type optimal control field obtained for a parametric nonlinear oscillator defined in Eq.~(\ref{anharmonic}), as compared with
   that of an approximate FFAD field obtained in the absence of
    the quartic term in the system Hamiltonian. The difference between the solid and dashed lines indicates
    the impact of the nonlinearity.      }\label{Fig4}
\end{figure}

\begin{figure*}
  \centering
  \includegraphics[width=13cm]{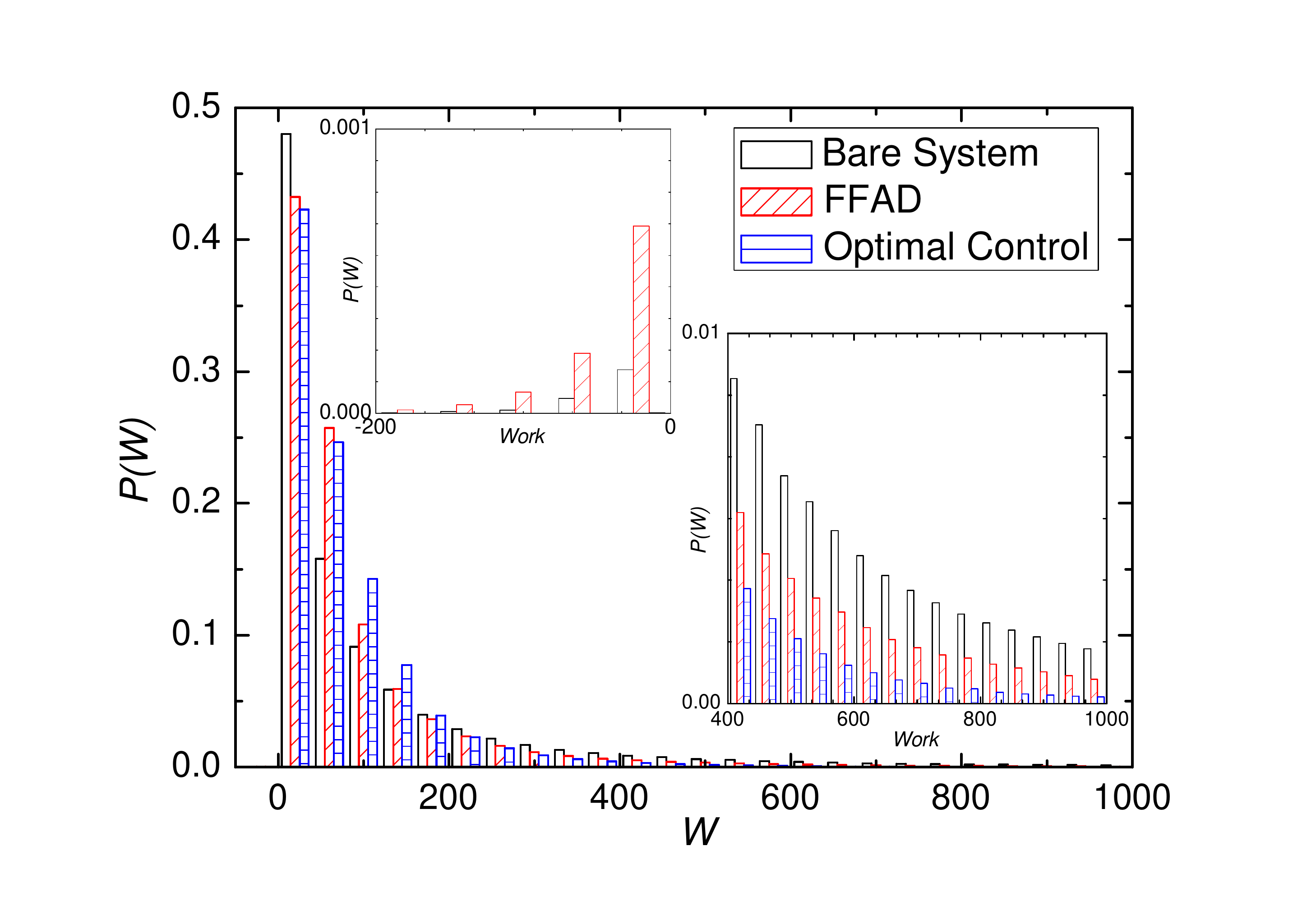}\\
  \caption{(color online)  Probability density of work distribution, in the absence of presence of
  different types of control fields. The results are obtained along with those presented in Table~\ref{performance2}, with all the system parameters of a parametric nonlinear oscillator
  the same as that described in  Table~\ref{performance2}.  Note that the optimal control field has suppressed
  a long tail of the work distribution and has also almost
 completely suppressed negative work values.}\label{Fig3}
\end{figure*}

To enhance the nonlinear effect, we next consider a case with a much higher temperature, i.e., $\beta=0.01$ (still
with $\varepsilon=1000$, $\omega_0=10.0$, and $\tau=0.001$).   The time-dependence of the found
optimal control field is presented in Fig.~\ref{Fig3}, which is seen to be very different from that of approximate FFAD.
To understand the details better, we also count the number of trajectories that may give a negative work output. As shown previously, for the protocol here the work would be always positive if the nonlinear term were not introduced.  So the presence of negative work values in the bare system
 does indicate the presence of nonlinearity.  Detailed computational results are shown in Table~\ref{performance2}.   It is seen that
the performance of optimal control is much better than that achieved by approximate FFAD.  While the optimal control field has essentially suppressed almost all negative work values (but one out of one million), approximate FFAD increases probabilities of negative work.  The reason why our optimal control approach is so effective in removing negative work values is quite simple.  For cases with a positive $\Delta F$,  the value of $e^{-\beta W}$ with a negative $W$ would be too drastically larger than its ``target" value $e^{-\beta \Delta F}$, so it will be rejected by our optimization algorithm as extreme or rare values of $e^{-\beta W}$. Note also that the variance in work under the optimal control field is less than one quarter of that for the bare system, and less than half of that obtained with approximate FFAD.  The mean work $\langle W\rangle$ under the optimal control field is also significantly smaller than
that in the bare system or in the FFAD case.   Because $\langle W\rangle-\Delta F$ is an interesting quantity called the dissipated work, a decrease in $\langle W\rangle$ indicates less dissipated work.
We also show in Fig.~\ref{Fig3} the work probability distribution $P(W)$.
There it is seen that the optimal control field most effectively suppresses very large or very small work values, in addition to
an almost complete removal of negative work values.
\begin{table*}[htb]
\caption{ The performance of work
fluctuation suppression in the absence or presence of several different control fields, mainly
characterized by the variance in work $W$ and the variance in $e^{-\beta W}$, using $10^6$ trajectories.
The system is a parametric nonlinear oscillator defined in Eq.~(\ref{anharmonic}),
and the time dependence of $\omega(t)$ is still given in Eq.~(\ref{frequency}),
with $\omega(0)=10$, $\omega(\tau)=10\sqrt{3}$, and $\tau=0.001$. The nonlinear parameter $\epsilon=1000$ and
the inverse temperature is set to be $\beta=0.01$ to enhance anharmonic effects.}
\vskip 1em
\centering
\begin{tabular}{ccccc}\hline\hline
{\quad}Process{\quad} & ${\quad}\sigma (e^{-\beta W}){\quad}$ & ${\quad}\langle W \rangle{\quad}$ & ${\quad}\sigma (W){\quad}$ & {\quad}Probability{\;}of{\;}negative{\;}work{\quad}\\[0.5ex]
\hline
bare{\;}system & $0.3440$ & $147.66$ & $428.54$ & $227\times10^{-6}$\\
\hline
FFAD{\;}control & $0.2905$ & $101.72$ & $249.57$ & $1002\times10^{-6}$ \\
\hline
optimal{\;}control{\;}of $pq$-type & $0.2701$ & $78.112$ & $102.51$ & $1\times10^{-6}$ \\
\hline\hline
\end{tabular}
\label{performance2}
\end{table*}

\section{Optimal Control of Work Fluctuations in Systems with Unknown Parameters}
The flexibility in optimal control and early studies of feedback control motivated us to
ask whether OCT can be used to suppress work fluctuations in those systems that have unknown system parameters. At the first thought, since with unknown system parameters, it is impossible to calculate $\Delta F$ and thus
our optimal control approach seems not applicable.  However, we show in this section that it is possible to
refine the control target function on the go, by iteratively guessing the target function through Jarzynski's equality.  This possibility should be good news for the application of Jarzynski's equality itself, as we can now indeed predict $\Delta F$ from non-equilibrium work values, with suppressed work fluctuations and hence better performance in reaching the correct $\Delta F$ based on a finite number of trajectories.

\begin{figure}
  \centering
  \includegraphics[width=8.4cm]{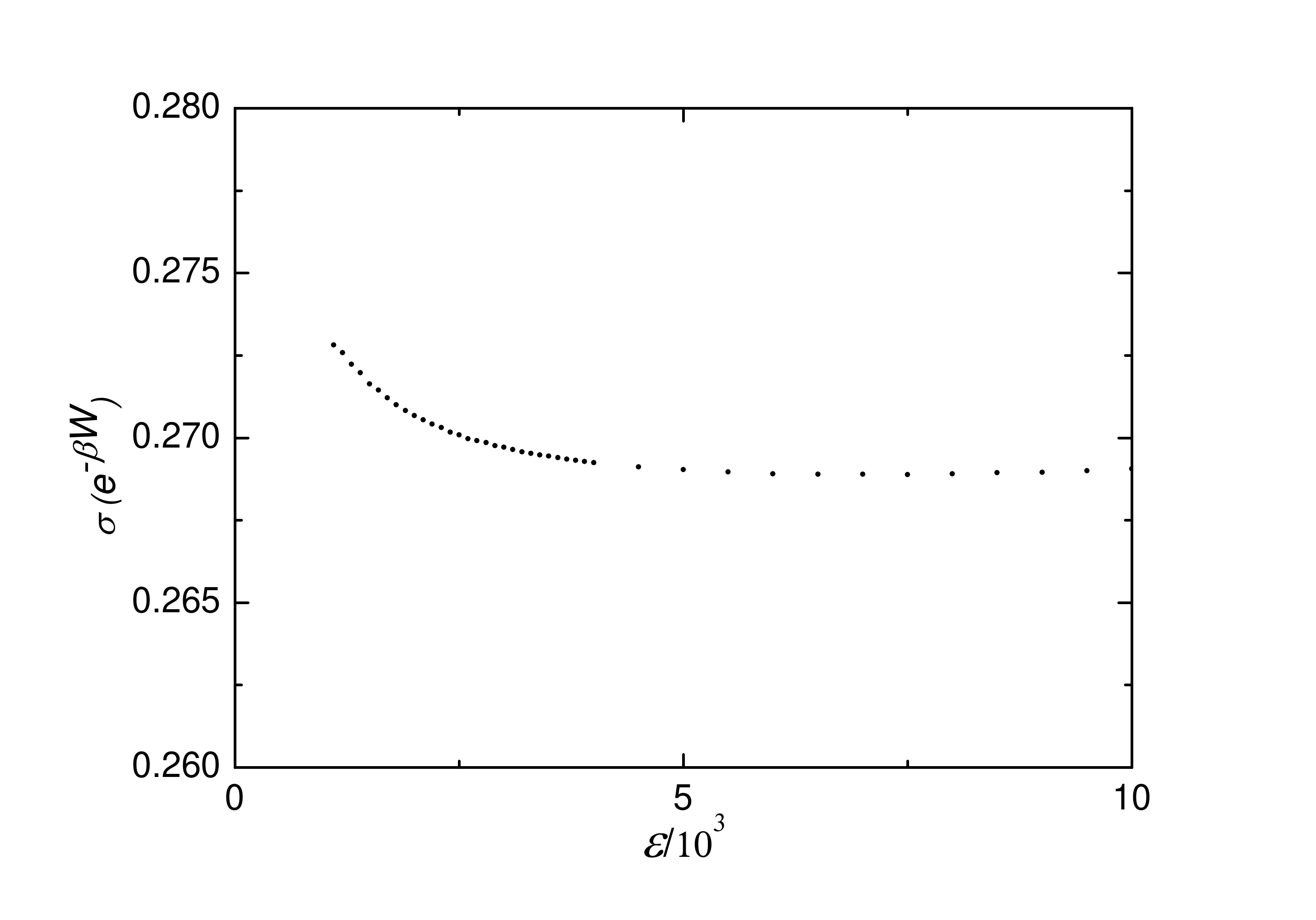}\\
  \caption{Variance of $ e^{-\beta W}$ obtained from optimal control applied to a parametric nonlinear oscillator described in Eq.~(\ref{anharmonic}), as a function of incorrectly preassumed values of the nonlinear parameter $\varepsilon$.  The actual nonlinear parameter $\varepsilon_0=1460$. $\omega(t)$ is still given by
  Eq.~(\ref{frequency}). Other system parameters are $\omega_0=10$, $\tau=0.001$, and $\beta=0.01$.   It is seen that despite wrong values of $\varepsilon$ are used in searching for optimal control fields, the obtained variance of $ e^{-\beta W}$ does not change significantly. }\label{Fig5}
\end{figure}

Let us first examine how our OCT framework depends on the knowledge of the bare system Hamiltonian. The first two relations in Eq.~(\ref{itteration}) require information of the initial thermal distribution and the target function. In other words, the full knowledge of the Hamiltonian in the beginning and at the end of
the protocol is needed there.  The third and fourth relations in Eq.~(\ref{itteration}) describe continuous time evolution
under the total Hamiltonian including the control field. However, if the protocol is fast, then the main component of the total Hamiltonian can be the control field, so the bare system Hamiltonian may not play an important role there.

To have some quantitative ideas, we consider again the anharmonic Hamiltonian in Eq.~(\ref{anharmonic}), i.e.,
$H_0(t)=\frac{p^2}{2m}+\frac{1}{2} m\omega^2(t) q^2+\varepsilon_0 q^4 \sin(\pi \frac{t}{\tau})$, with $\omega_0=10$ and $\tau=0.001$.
For this designed case, the quartic term vanishes at $t=0$ and $t=\tau$,
$\Delta F$ and hence the target function in our OCT is still fully known.
Next we set $\varepsilon_0=1460$, but in our construction and implementation of OCT we do not use
this piece of knowledge.  Instead, we use some wrong values of $\epsilon$ when computationally search for
the OCT field. The knowledge of $\varepsilon_0=1460$ is used for checking only,  {\it after} we have obtained the OCT field and start to look into the actual work fluctuations with the control field thus obtained.

As shown in Fig.~\ref{Fig5}, with $\varepsilon$ used in our optimal control algorithm
varying from $1000$ to $10000$,
the fluctuations in $e^{-\beta W}$ is not changing significantly.  That is,  for a wide range of incorrectly assumed values of $\varepsilon$, the associated optimal control field can still effectively suppress work fluctuations, with $\sigma\left(e^{-\beta W}\right) $ ranging from $0.269$ to $0.273$.  Additional numerical investigations of the evolving trajectories further indicate that the total time duration is too short for $H_0$ to play a role, thus confirming our qualitative insights above.

This computational example is enlightening, but we assume there that
the (bare) system Hamiltonian is fully known at two boundary times $t=0$ and $t=\tau$.
Since in such situations $\Delta F$ can be exactly calculated from the bare system Hamiltonian,
we are still one step away from {\it predicting} $\Delta F$ using the work values in a nonequilirium protocol.
So our final question is the following, if some parameter in the (bare) system Hamiltonian is indeed unknown to us, how to construct the optimal control target function, suppress the work fluctuations, and eventually predict $\Delta F$?

Borrowing the idea from feedback optimal control theory \cite{Judson.92.PRL}, we aim to
propose
a useful computational feedback procedure as illustrated in Fig.~\ref{Fig6}.  It consists of the following steps:
\begin{figure}
  \centering
  \includegraphics[width=8.4cm]{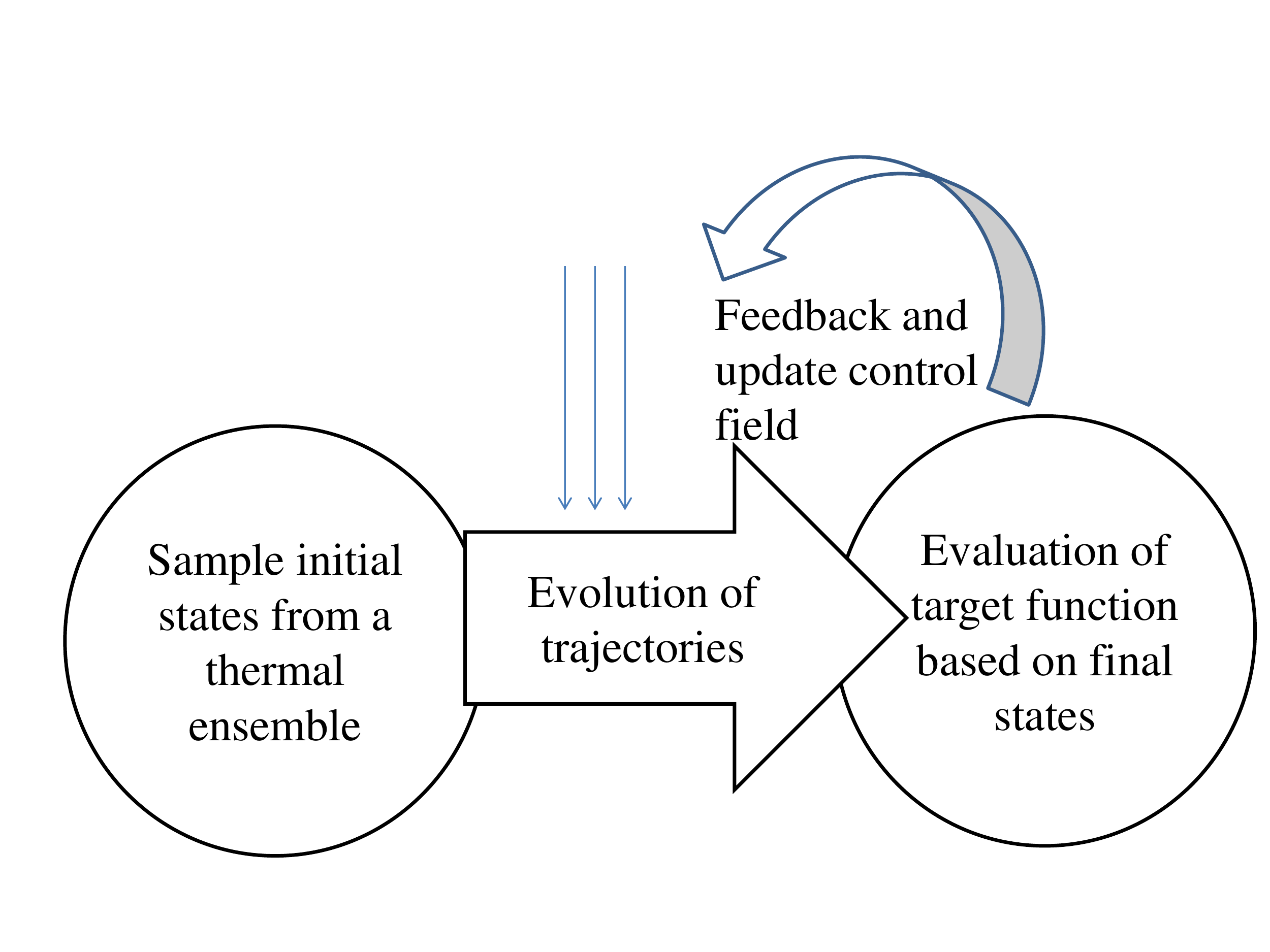}\\
  \caption{Procedure of executing a feedback loop to construct OCT fields iteratively, in order to apply
  OCT to systems with unknown system parameters.  The feedback is to refine the control target function based on Jarzynski's equality,  by use of Jarzynski's equality itself to guess the free energy difference.}\label{Fig6}
\end{figure}
\begin{itemize}
  \item[(i)] A certain small  number of initial states are first sampled according to system's thermal distribution, and then
  evolved in accord with a guessed control field.
  \item[(ii)] Based on the previous step, a rough estimate of $\langle e^ {-\beta W}\rangle$ is obtained to yield $ e^ {-\beta \Delta F}$,   which is then used to yield/update the control target function.
  \item[(iii)] To optimize the control field based on Eq.~(\ref{itteration}),
  one may neglect the effect of the bare system Hamiltonian or preassume some wrong parameter values
  because the evolution is mainly dictated by the control field during a rapid work protocol.
  \item[(iv)] The control field is then updated by the output from an optimal control algorithm and then all the previous steps are repeated until some convergence threshold is met.
\end{itemize}

\begin{table*}[htb]
\caption{
The performance of work
fluctuation suppression in the absence or presence of several different control fields, mainly
characterized by the variance in work $W$ and the variance in $e^{-\beta W}$, using $10^6$ trajectories.
The system is a parametric nonlinear oscillator defined in Eq.~(\ref{anharmonic2}),
whose quartic term does not vanish at the end of the protocol.  The coefficient of the quartic term $\varepsilon=1460$ is never used when searching for the control field via feedback OCT.  The control field is obtained by five iterations of the feedback loop illustrated in Fig.~\ref{Fig6}.
Other system parameters are $\omega_0=10$, $\tau=0.001$, and $\beta=0.01$, $\omega(t)$ is given by
  Eq.~(\ref{frequency}).}
\vskip 1em
\centering
\begin{tabular}{ccccc}\hline\hline
{\quad}Process{\quad} & ${\quad}\langle e^{-\beta W} \rangle{\quad}$ & ${\quad}\sigma (e^{-\beta W}){\quad}$ & ${\quad}\langle W \rangle{\quad}$ & ${\quad} \sigma (W) {\quad}$\\[0.5ex]
\hline
bare{\;}system & $0.3269$ & $0.4031$ & $4453$ & $14288$\\
\hline
Feedback OCT{\;}$(\varepsilon=1000)$ & $0.3272$ & $0.3370$ & $631.0$ & $1787$\\
\hline
Feedback OCT{\;}$(\varepsilon=2000)$ & $0.3273$ & $0.3366$ & $528.4$ & $1409$\\
\hline
Feedback OCT{\;}$(\varepsilon=3000)$ & $0.3273$ & $0.3367$ & $479.7$ & $1220$\\
\hline\hline
\end{tabular}
\label{performance3}
\end{table*}

\begin{figure}
  \centering
  \includegraphics[width=9.4cm]{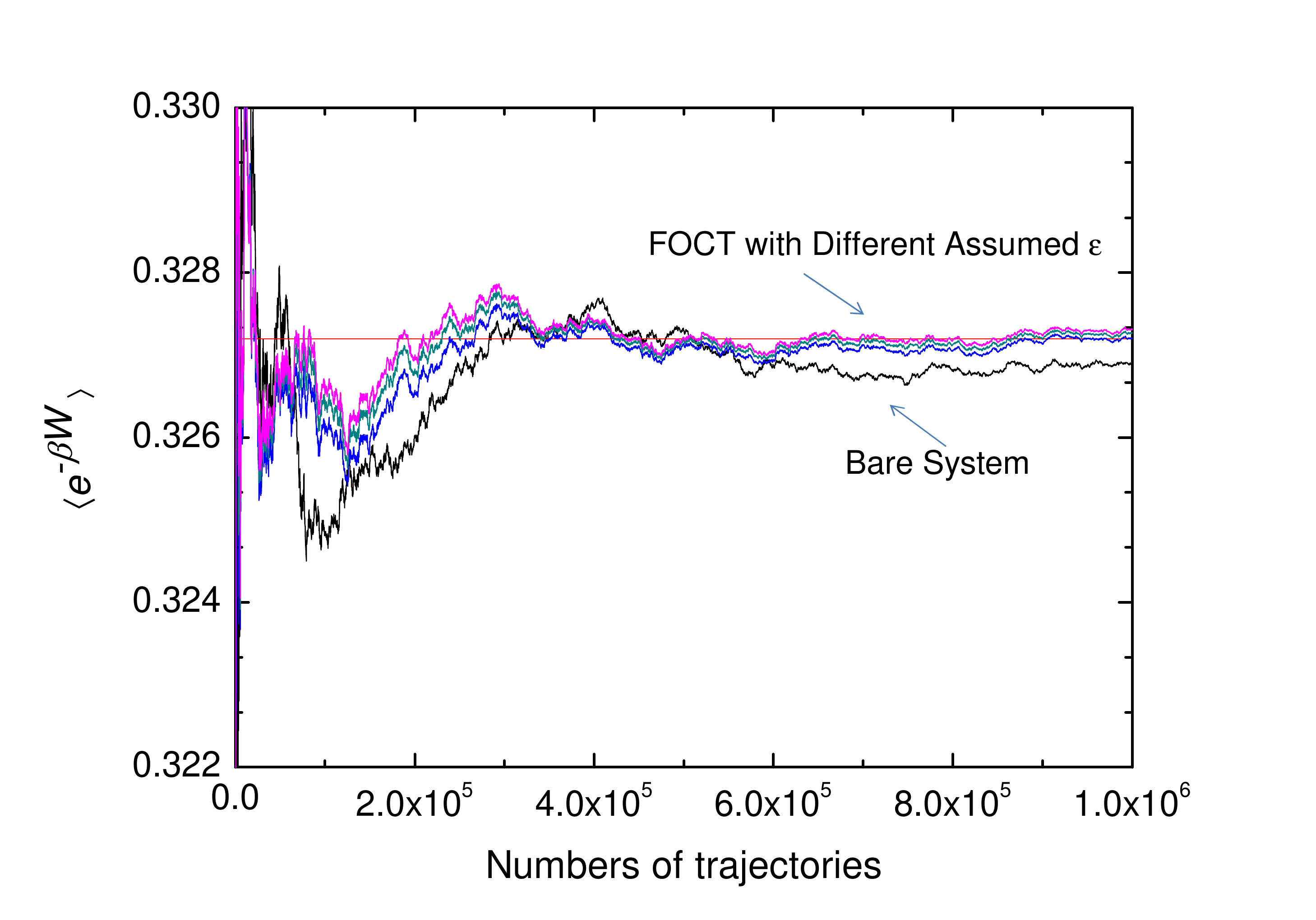}\\
  \caption{(color online) The convergence of $\langle e^{-\beta W} \rangle$ towards the theoretical value (indicated by the horizontal line) versus the number of trajectories used in the simulations. The system
  and all the system parameters are all specified in Table~\ref{performance3}.  It is seen that the three cases with optimal control fields obtained from feedback mechanism can yield better results, despite that the actual value of the nonlinear parameter is not needed in constructing the control fields.
   }\label{Fig8}
\end{figure}

To demonstrate our strategy we consider the following system
\begin{equation}\label{anharmonic2}
  H_0(t)=\frac{p^2}{2m}+\frac{1}{2} m\omega(t)^2 q^2+\varepsilon_0 q^4 \sin\left(\frac{\pi}{2} \frac{t}{\tau}\right).
\end{equation}
Note that in this case, the quartic term is not vanishing at $t=\tau$, with $\varepsilon_0=1460$ assumed to be ``unknown" when we seek the optimal control field. That is, we will not use this value to construct our control field. Instead, we use some pre-assumed wrong values of $\epsilon$ when solving Eq.~(\ref{itteration}).  We still set the quartic term to be zero at $t=0$ for the convenience
in initial state sampling.  In the first iteration of our numerical experiment, we use a null control field to start with.  We update the target function and the optimal control field based on $20, 000$ trajectories only. Remarkably,
the optimal control field can already converge well after only five iterations of the above four steps.
In particular, we compare in Table \ref{performance3} the results from $10^6$ trajectories,
in terms of $\langle e^{-\beta W}\rangle$, $\sigma(W)$, $\langle W\rangle$, and
 $\sigma\left(e^{-\beta W}\right)$.
In obtaining the numerical results
 we have used three different preassumed values of $\varepsilon$ (all are much different from the real value).
As seen from the third column of Table \ref{performance3}, for all three cases,
the fluctuations in $e^{-\beta W}$ are effectively well suppressed  (as compared with the bare case without a control field). The ensemble-average $\langle e^{-\beta W}\rangle$ in all the three cases are also close to the true theoretical value 0.3272 (obtained by numerically computing the partition function of $H_0$ with $\varepsilon_0=1460$).  Interestingly, the value of $\langle e^{-\beta W}\rangle$ for the bare system case (second column, second row of  Table \ref{performance3})
is still slightly away from this theoretical value.  This suggests that in the bare system case
a high-quality convergence towards Jarzynski's equality has not been achieved with $10^6$ trajectories.  Thus, the presence of a control field suppressing work fluctuations is seen to have, albeit slightly, enhanced the convergence of the simulation towards Jarzynski's equality. To see this more clearly,
we show in Fig.~\ref{Fig8} how the numerically obtained average value of $e^{-\beta W}$ gradually converges
towards the theoretical value $e^{-\beta \Delta F}$, as the number of trajectories increases to $10^6$.
The horizontal line in Fig.~\ref{Fig8} represents the true theoretical value. It is seen that all the three cases of feedback control have converged to the true theoretical value with about $6\times 10^5$ trajectories, but the bare system case still has a non-negligible shift from the theoretical result.

Returning to Table \ref{performance3}, from the last two columns it is also seen that
the mean work  $\langle W \rangle$ (hence also the dissipated work)
 as well as the variance of $W$ is suppressed by the optimal control field by about one order of magnitude.  This significant control over the work output and its fluctuations
is achieved even though part of the system parameters is unknown to us!   Interestingly, this does not mean
that the variance of $e^{-\beta W}$ (see second column of Table \ref{performance3})
will be also significantly reduced by the control field.  Qualitatively, note that a higher temperature  induces a wider initial probability distribution and hence larger thermal fluctuations in work values, but on the other hand, when calculating $e^{-\beta W}$ the larger work values are still scaled down by the inverse temperature $\beta$.  As such,  from the explicit
computational example here it is learned that one should not underestimate the implications of a seemingly ``small" suppression in the variance of $e^{-\beta W}$.


\section{Concluding Remarks}

In this paper we have proposed an optimal control approach aiming at the suppression of work fluctuations
associated with a protocol applied to a thermal ensemble.  Our approach is original in the sense that
we directly exploits Jarzynski's equality in defining our control target function.  Indeed, the control target function is constructed to bias against the deviation of individual values of $e^{-\beta W}$ from their ensemble average  $\langle e^{-\beta W}\rangle = e^{-\beta \Delta F}$, where $\Delta F$ is the free energy difference.  This approach is shown to be very effective. In the case of a parametric oscillator, the performance of our optimal control approach is simply as good as previous methods based on accelerated adiabatic processes. More importantly, our optimal control approach can equally apply to rather arbitrary nonlinear systems.  We also note that a recent study \cite{Lu.14.PRA} considered  a somewhat related optimal control approach for nonlinear systems, but assuming weak anharmonic potential to validate a perturbative treatment.  Our approach can however be applied to systems with strong nonlinear effects, as we have shown through nonlinear oscillator systems with a large quartic term.  One might also think that since $\Delta F$ is needed in the construction of a target function, we may have to know the full information of the system Hamiltonian.  Through simple numerical experiments, we showed that this intuition is not true, because a feedback mechanism can help us to refine the control target function on the fly.  As such, we claim that the suppression of work fluctuations via a control field is possible even when we do not have full knowledge of the system.

We believe that this study is just a motivator and a starting point along the general issue of controlling and manipulating nonequilibrium thermodynamic properties.  Many follow up studies are on the way.  For example,
it is of immediate interest to extend this study to quantum systems in order to account for quantum effects and quantum fluctuations.  So far we have assumed sufficiently rapid work protocols during which system-bath interaction is neglected. In the future it is necessary to extend our approach here to open systems.  To that direction we note an early study considering classical overdamped motion under a special type of optimization \cite{Schmiedl.07.PRL}.

\section{Ackowledgements}

We are grateful to  Peter Hanggi, Dario Poletti, and Manas Mukherjee for interesting discussions.

\newpage

\section{Appendix: The Iteration Procedure of OCT}
In this section we discuss numerical iteration details in seeking optimal control fields from Eq.~(\ref{itteration}).  The procedure is rather standard in OCT \cite{Shi.90.JCP}. First, we divide the relations in Eq.~(\ref{itteration}) together with Hamilton's equations of motion into four parts.
\begin{equation}\label{itteration1}
  \left\{
  \begin{split}
      & \frac{e^{-{\beta} H_0(q_0,p_0,0)}}{Z_0}{\,} \frac{\partial Q}{\partial q_\tau}+{\lambda_\tau}=0, \\
      & \frac{e^{-{\beta} H_0(q_0,p_0,0)}}{Z_0}{\,} \frac{\partial Q}{\partial p_\tau}+{l_\tau}=0;
  \end{split}
  \right.
\end{equation}
\begin{equation}\label{itteration2}
  \left\{
  \begin{split}
      & -\dot{\lambda}(q,p,t)+l(q,p,t) {\frac{\partial^{2}{H}}{\partial{q^2}}}-{\lambda(q,p,t)} {\frac{\partial^{2}{H}}{\partial{p} \partial {q}}}=0,\\
      & -\dot{\l}(q,p,t)+l(q,p,t) {\frac{\partial^{2}{H}}{\partial{p} \partial {q}}}-{\lambda(q,p,t)} {\frac{\partial^{2}{H}}{\partial{p^2}}}=0;
  \end{split}
  \right.
\end{equation}
\begin{equation}\label{itteration3}
       {\int_\Gamma} \left[l(q,p,t){\frac{\partial^{2}{H}}{\partial{q} \partial {A}}}-\lambda(q,p,t){\frac{\partial^{2}{H}}{\partial{p} \partial {A}}}\right]dq_0dp_0+\kappa A(t)=0;
\end{equation}
\begin{equation}
  \left\{
  \begin{split}
       \dot{q} &  = \frac{\partial H}{\partial p}, \\
       \dot{p} &  = -\frac{\partial H}{\partial q}.
  \end{split}
  \right.
  \label{motion1}
\end{equation}
 Next, in numerical calculations, we can consider a sufficiently large but finite
 phase space area, by considering the initial thermal distribution in Eq.~(\ref{Gibbs}).
 For each initial phase space point $(p_0,q_0)$ sampled, one integrates Hamilton's equation from $t=0$ to $t=\tau$, using a guessed field time dependence $A(t)$. Then the final values of the Lagrange multipliers $\lambda(q,p,\tau)$ and $l(q,p,\tau)$ can be computed from Eq.~(\ref{itteration1}. With this, Eq.~(\ref{itteration2}) can be  integrated backwards. Finally, with $[p(t),q(t)]$, $\lambda(q,p,t)$, and $l(q,p,t)$ all solved numerically, the field time dependence $A(t)$ for the next iteration can be obtained by referring to Eq.~(\ref{itteration3}). This iterative procedure is repeated until certain convergence criteria are satisfied. In our implementation of this procedure
 we simply set the starting $A(t)$ to be zero. At the end of each iteration, we update the field time dependence for the next iteration as follows:
\begin{eqnarray}\label{newcontrol}
 &&\left\{A(t)-\frac{1}{\kappa}{\int_\Gamma} \left[l(q,p,t){\frac{\partial^{2}{H}}{\partial{q} \partial {A}}}-\lambda(q,p,t){\frac{\partial^{2}{H}}{\partial{p} \partial {A}}}\right]dq_0dp_0\right\} \nonumber \\
 && \rightarrow A(t).
\end{eqnarray}
Basically, the iteration equation means that a difference between final state and target state is calculated after each iteration, and then a modification of the control field is made.


\begin{thebibliography}{99}%
\bibitem{Jarzynski.97.PRL} C. Jarzynski, Phys. Rev. Lett. {\bf 78}, 2690 (1997).
\bibitem{Jarzynski.97.PRE} C. Jarzynski, Phys. Rev. E {\bf 56}, 5018 (1997).
\bibitem{crooks} G. E. Crooks, \pre{\bf 60}, 2721 (1999).
\bibitem{Campisi.11.RMP} M. Campisi, P. H\"{a}nggi, and P. Talkner, Rev. Mod. Phys. {\bf 83}, 771 (2011).
\bibitem{Abah.12.PRL} O. Abah, J. Ro{\ss}nagel, G. Jacob, S. Deffner, F. Schmidt-Kaler, K. Singer, and E. Lutz, Phys.
Rev. Lett. {\bf 109}, 203006 (2012).
\bibitem{Bergenfeldt.14.PRL} C. Bergenfeldt, P. Samuelsson, B. Sothmann, C. Flindt, and M. B¨¹ttiker, Phys. Rev. Lett.
{\bf 112}, 076803 (2014).
\bibitem{Zhang.14.PRL} K. Zhang, F. Bariani, and P. Meystre, Phys. Rev. Lett. {\bf 112}, 150602 (2014).
\bibitem{Deng.13.PRE} J. Deng, Q. Wang, Z. Liu, P. H\"{a}nggi, and J. B. Gong, Phys. Rev. E {\bf 88}, 062122 (2013).
\bibitem{dario}Y. Zheng and D. Poletti, \pre\ {\bf 90}, 012145 (2014).
\bibitem{campoarxiv} A. del Campo, J. Goold, and M. Paternostro, arXiv:1305.3223.
\bibitem{Shi.90.JCP} S. Shi and H. Rabitz, J. Chem. Phys. {\bf 92}, 364 (1990).
\bibitem{Peirce.88.PRA} A. P. Peirce, M. A. Dahleh, and H. Rabitz, Phys. Rev. A {\bf 37}, 4950 (1988).
\bibitem{Shi.91.CPC} S. Shi and H. Rabitz, Comput. Phys. Commun. {\bf 63}, 71 (1991).
\bibitem{Sundermann.99.JCP} K. Sundermann and R. de Vivie-Riedle, J. Chem. Phys. {\bf 110}, 1896 (1999).
\bibitem{Ohtsuki.04.JCP} Y. Ohtsuki, G. Turinici, and H. Rabitz, J. Chem. Phys. {\bf 120}, 5509 (2004).
\bibitem{Demirplak.08.JCP} M. Demirplak and S. A. Rice, J. Chem. Phys. {\bf 129}, 154111 (2008).
\bibitem{Demirplak.05.JPCB} M. Demirplak and S. A. Rice, J. Phys. Chem. B {\bf 109}, 6838 (2005).
\bibitem{Demirplak.03.JPCA} M. Demirplak and S. A. Rice, J. Phys. Chem. A {\bf 107}, 9937 (2003).
\bibitem{Berry.09.JPAMT} M. V. Berry, J. Phys. A: Math. Theor. {\bf 42}, 365303 (2009).
\bibitem{Masuda.10.PRSLSA} S. Masuda and K. Nakamura, Proc. R. Soc. London Ser. A {\bf 466}, 1135 (2010).
\bibitem{Chen.10.PRL} X. Chen, I. Lizuain, A. Ruschhaupt, D. Gu\'{e}ry-Odelin, and J. G. Muga, Phys. Rev. Lett.
{\bf 105}, 123003 (2010).
\bibitem{Chen.10.PRLa} X. Chen, A. Ruschhaupt, S. Schmidt, A. del Campo, D. Gu\'{e}ry-Odelin, and J. G. Muga,
Phys. Rev. Lett. {\bf 104}, 063002 (2010).
\bibitem{Choi.11.PRA} S. Choi, R. Onofrio, and B. Sundaram, Phys. Rev. A {\bf 84}, 051601 (2011).
\bibitem{Schaff.11.NJP} J. F. Schaff, P. Capuzzi, G. Labeyrie, and P. Vignolo, New J. Phys. {\bf 13}, 113017 (2011).
\bibitem{Bason.12.NP} M. G. Bason, M. Viteau, N. Malossi, P. Huillery, E. Arimondo, D. Ciampini, R. Fazio, V. Giovannetti, R. Mannella, and O. Morsch, Nat. Phys. {\bf 8}, 147 (2012).
\bibitem{Zhang.13.PRL} J. Zhang, J. H. Shim, I. Niemeyer, T. Taniguchi, T. Teraji, H. Abe, S. Onoda, T. Yamamoto,
T. Ohshima, J. Isoya, and D. Suter, Phys. Rev. Lett. {\bf 110}, 240501 (2013).
\bibitem{Jarzynski.13.PRA} C. Jarzynski, Phys. Rev. A {\bf 88}, 040101 (2013).
\bibitem{Campo.11.PRA} A. del Campo, Phys. Rev. A {\bf 84}, 031606 (2011).
\bibitem{Deffner.14.PRX} S. Deffner, C. Jarzynski, and A. del Campo, Phys. Rev. X {\bf 4}, 021013 (2014).
\bibitem{mugareview} E. Torrontegui, S. Ibanez, S. Mart\'{i} nez-Garaot, M. Modugno, A. del Campo, D. Gu\'{e}ry-Odelin, A. Ruschhaupt, X. Chen, and J. G. Muga, Adv. At. Mol. Opt. Phys. 62, 117-169 (2013).
\bibitem{Tu.14.PRE} Z. C. Tu, Phys. Rev. E {\bf 89}, 052148 (2014).
\bibitem{Campo.13.PRL} A. del Campo, Phys. Rev. Lett. {\bf 111}, 100502 (2013).
\bibitem{Jarzynski.07.CRP} C. Jarzynski, C. R. Phys. {\bf 8}, 495 (2007).
\bibitem{Rohringer.13.a1} W. Rohringer, D. Fischer, F. Steiner, I. E. Mazets, J. Schmiedmayer, and M. Trupke, arXiv
1312.5948 (2013).
\bibitem{Judson.92.PRL} R. S. Judson and H. Rabitz, Phys. Rev. Lett. {\bf 68}, 1500 (1992).
\bibitem{Lu.14.PRA} X. J. Lu, X. Chen, J. Alonso, and J. G. Muga, Phys. Rev. A {\bf 89}, 023627 (2014).
\bibitem{Schmiedl.07.PRL} T. Schmiedl and U. Seifert, Phys. Rev. Lett. {\bf 98}, 108301 (2007).
\end{thebibliography}
\end{document}